\RequirePackage{fix-cm}
\documentclass[twocolumn]{svjour3}          % twocolumn
\smartqed  % flush right qed marks, e.g. at end of proof
\usepackage{graphicx}
\usepackage{url}
\usepackage{physics}
\usepackage{xcolor}
\usepackage{amsfonts}

\usepackage[colorlinks=true, allcolors=blue]{hyperref}

\usepackage{amsmath}
\usepackage{smartdiagram}
\usepackage{tikz}
\usesmartdiagramlibrary{additions}

\tikzstyle{mynode}=[thick,draw=blue,fill=blue!20,circle,minimum size=22]
\usetikzlibrary{calc}
\usetikzlibrary{arrows}

\newcommand{\icol}[1]{% inline column vector
  \left(\begin{smallmatrix}#1\end{smallmatrix}\right)%
}

\usepackage{tikz}
\usetikzlibrary{quantikz}

\def\makeheadbox{\relax}

\begin{document}
\title{Quantum Artificial Intelligence: A Brief Survey
%\title{Quantum Artificial Intelligence
%\title{Quantum Artificial Intelligence: Status and Challenges
%\title{Quantum Artificial Intelligence: An Overview

}


\author{ Matthias Klusch \and Jörg Lässig \and Daniel Müssig \and \\
         Antonio Macaluso \and Frank K. Wilhelm
}

\authorrunning{Klusch et al.} % if too long for running head

\institute{Matthias Klusch \at
              German Research Center for Artificial Intelligence GmbH (DFKI), Saarbruecken, Germany\\
              \email{matthias.klusch@dfki.de}           %  \\
           \and
           Jörg Lässig \at
              University of Applied Sciences Zittau/Görlitz, Görlitz, Germany\\
              \email{jlaessig@hszg.de}
           \and
           Daniel Müssig \at
              Fraunhofer Institute of Optronics, System Technologies and Image Exploitation - Advanced System Technology (IOSB-AST), Görlitz, Germany\\
              \email{daniel.muessig@iosb-ast.fraunhofer.de}
            \and
           Antonio Macaluso \at
              German Research Center for Artificial Intelligence GmbH (DFKI), Saarbruecken, Germany\\        
              \email{antonio.macaluso@dfki.de}           %  \\
            \and
           Frank K. Wilhelm \at
           Forschungszentrum Juelich GmbH (FZJ), Jülich, Germany
           \email{f.wilhelm-mauch@fz-juelich.de}
}

\date{ }


\maketitle

\begin{abstract}
Quantum Artificial Intelligence (QAI) is the intersection of quantum computing and AI,
a technological synergy with expected significant benefits for both. 
In this paper, we provide a brief overview of what has been achieved in QAI so far
and point to some open questions for future research.
In particular, we summarize some major key findings on the feasability and the potential of 
using quantum computing for solving computationally hard problems in various subfields 
of AI, and vice versa, the leveraging of AI methods for building and operating quantum 
computing devices.  

%Include keywords, PACS and mathematical subject classification numbers as needed.
%\PACS{PACS code1 \and PACS code2 \and more}
\end{abstract}

% === 1 =====================================================
\section{Introduction}%
\label{sec:intro}%

\noindent
It is known that quantum computing can simulate and even go beyond classical computing in terms 
of computational speedup in theory \cite{A99,NC01,RP14} for quite some time. But initial versions of real 
quantum computing hardware and frameworks for quantum programming became available only in about 
the past decade. Even the Quantum Internet of networked quantum computers and with secure quantum 
communication channels, long time considered as mere science fiction, is on its way with early stage 
prototypes available \cite{C24,QIA,QCN21}. 
On the other hand, artificial intelligence (AI) \cite{RN16} is commonly considered as one of the most 
disruptive key technologies of our time for industry and business, our private and social life, 
notwithstanding the challenges of its future, trustworthy and controlled use for the benefit of the 
people affected by it. 

% FIGURE QAI - one-column wide figure
\begin{figure}
\begin{center}
\includegraphics[width=0.9\columnwidth]{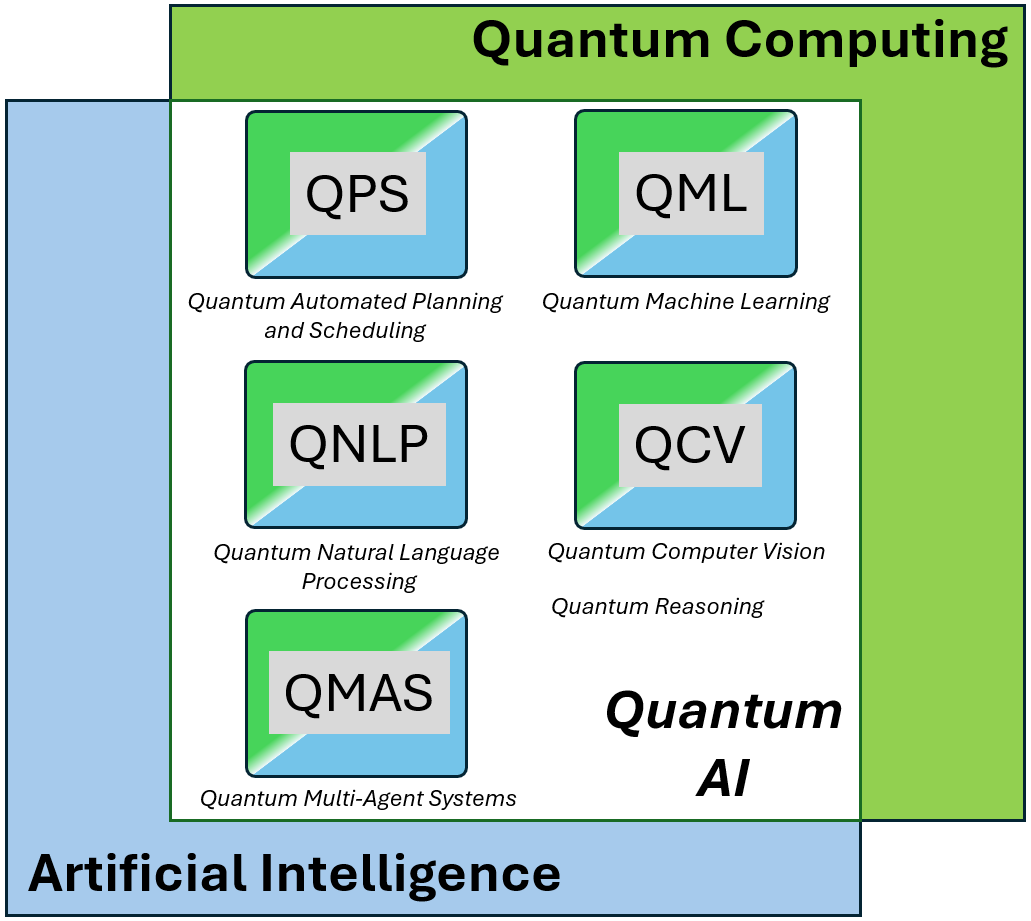}
\caption{Quantum AI (QAI) as intersection of quantum computing and AI with subfields in relation to AI 
each covering both directions.}
\label{fig:1}     
\end{center}
\end{figure}

\noindent
{\em Quantum Artificial Intelligence} (QAI, Quantum AI) is the intersection of both technologies 
(cf. Fig. 1) and concerned with the investigation of the feasability and the potential of leveraging 
quantum computing for AI, and vice versa, AI for quantum computing \cite{R23}. 
While quantum machine learning \cite{B+20,B+22} is currently the most popular application 
\cite{S23,D23,D23b,MLV23}, quantum AI goes much beyond covering more 
subfields of AI \cite{RN16,GRS20}, 
such as quantum 
reasoning \cite{C+23,CGL03,BvN1936}, quantum automated planning and scheduling (QPS), 
quantum natural language processing (QNLP), quantum computer vision (QCV), and quantum agents 
and multi-agent systems (QMAS). 
Notably, each of these QAI subfields covers research and development in both directions; 
for example, QML refers to both the use of quantum computing for machine learning and vice versa.
In this respect, it would be premature to follow the current hype cycle around QML.\\
\noindent
Though QAI still is a nascent and inherently interdisciplinary research field, remarkable progress 
has been already made in the recent past. It has been shown, in part also experimentally on real 
quantum computers of the current NISQ (Noisy Intermediate-Scale Quantum) era, that tailored QAI 
algorithms could indeed make a positive difference in solving certain computationally hard problems 
such as for combinatorial optimization in AI applications compared to classical solutions. 
So far, the potential quantum utility of direct quantum or hybrid quantum-classical methods of QAI 
is under investigation for applications in diverse domains such as manufacturing, automated driving, 
transport and logistics, energy management, healthcare, finance, aerospace, climate and earth 
sciences, and pharmaceutical and chemical industries. 
In this regard, quantum AI use cases are concerned with, for example,
portfolio optimization in finance \cite{RS24}, 
traffic management \cite{Z+24},
capacitive vehicle routing \cite{F+19,I+19}, 
safe navigation of self-driving vehicles \cite{SMK23}, 
satellite network constellation \cite{V+24a} and mission planning \cite{R+23}, 
energy network management \cite{eon1,eon2},
job-shop scheduling in manufacturing \cite{SWGA23},
weather simulation and forecasting \cite{J+24,GJ24,O+24,TP23} and
simulation of materials and chemicals for drug discovery \cite{P+23,K+23,M+23}
and material design \cite{C+24,A+24}. 
On the other hand, there is active research on the use of QAI algorithms that leverage AI 
methods, in particular from machine learning, to address challenges of the building and operation 
of quantum computing devices. 
These research activities in both directions of QAI are in part also fueled and influenced by 
the made progress and still ongoing race in building ever more powerful quantum computing devices 
and network, and vice versa.\\
\noindent
Currently, the global market value of QAI applications on future quantum computing devices in 
general is estimated to be about eighteen billion USD by 2030 \cite{AMR24}, and three to five 
billion USD for the automotive industry by 2035 in particular \cite{RM23,BMP20}. This emphasizes that quantum AI has graduated from being a mere academic niche to a topic with a potential future beyond the current 
hype. However, it remains unclear when QAI methods and their applications, including those 
mentioned above, can be used and commercialized at large in practice, as this would require way 
more quantum computational resources and fault tolerance than current quantum computers have but 
future computing devices could provide \cite{KLW24}. \\
\noindent
Remarkably, as in AI, there are also discussions related to QAI on ethical issues of quantum computing 
such as those concerned with non-transparency of information processing in a 
\lq\lq quantum-box", or clashes of quantum privacy with security demands \cite{P23}, as well as on 
the role of quantum computing in neuroscience (quantum neuroscience). The latter includes 
%the advancement of computationally hard drug discovery such as for mental health, 
speculative multi-scale simulations of the human brain \cite{S+22} and explanation of some 
consciousness-related brain functions \cite{KP22}, though not of how the human brain produces 
thoughts \cite{L+06}, as well as a better representation of and inference means for 
certain psychological models in cognitive science (quantum cognition) \cite{PB22}. One should keep in mind, that these concerns apply in similar form to classical stochastic algorithms as well.  \\
\noindent
In this paper, we provide a first overview of selected methods, use cases and insights from 
research in the interdisciplinary field of QAI for both its directions, without any claim to 
completeness. In Section 2, we informally recall the basics of quantum computing from the computer 
science perspective only in very brief; readers who are roughly familiar with them can easily skip this 
section. Section 3 then summarizes selected key findings of research on quantum computing for AI in 
several subfields of QAI. The same is done for research on AI for quantum computing in Section 4 
before we conclude in Section 5.
% similar to the editorial introduction part: what is QAI?

% === 2 =====================================================
\section{Quantum Computing in a Nutshell}%
\label{sec:QC}%

\begin{figure}
    \centering
        \begin{tikzpicture}[line cap=round, line join=round, >=Triangle]
          \clip(-2.19,-2.49) rectangle (2.66,2.58);
          \draw [shift={(0,0)}, lightgray, fill, fill opacity=0.1] (0,0) -- (56.7:0.4) arc (56.7:90.:0.4) -- cycle;
          \draw [shift={(0,0)}, lightgray, fill, fill opacity=0.1] (0,0) -- (-135.7:0.4) arc (-135.7:-33.2:0.4) -- cycle;
          \draw(0,0) circle (2cm);
          \draw [rotate around={0.:(0.,0.)},dash pattern=on 3pt off 3pt] (0,0) ellipse (2cm and 0.9cm);
          \draw (0,0)-- (0.70,1.07);
          \draw [->] (0,0) -- (0,2);
          \draw [->] (0,0) -- (-0.81,-0.79);
          \draw [->] (0,0) -- (2,0);
          \draw [dotted] (0.7,1)-- (0.7,-0.46);
          \draw [dotted] (0,0)-- (0.7,-0.46);
          \draw (-0.08,-0.3) node[anchor=north west] {$\varphi$};
          \draw (0.01,0.9) node[anchor=north west] {$\theta$};
          \draw (-1.01,-0.92) node[anchor=north west] {$\mathbf {x}$};
          \draw (2.07,0.3) node[anchor=north west] {$\mathbf {y}$};
          \draw (-0.5,2.6) node[anchor=north west] {$\mathbf {z=|0\rangle}$};
          \draw (-0.4,-2) node[anchor=north west] {$-\mathbf {z=|1\rangle}$};
          \draw (0.4,1.65) node[anchor=north west] {$|\psi\rangle$};
          \scriptsize
          \draw [fill] (0,0) circle (1.5pt);
          \draw [fill] (0.7,1.1) circle (0.5pt);
        \end{tikzpicture}
     \caption{Representation of a quantum state \(|\psi\rangle\) on the Bloch sphere. The state is described by the angles \(\theta\) and \(\varphi\), where \(\theta\) defines the polar angle from the \(z\)-axis and \(\varphi\) defines the azimuthal angle in the \(xy\)-plane. The Bloch sphere provides a geometric representation of the pure states of a qubit, with \(|0\rangle\) and \(|1\rangle\) corresponding to the poles of the sphere.}
    \label{fig:block}
    \end{figure}
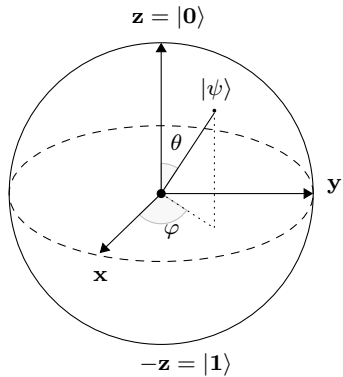

\noindent
Quantum computing \cite{NC01,RP14} harnesses the principles of quantum mechanics \cite{R21} to process information and perform computations, potentially surpassing the capabilities of classical computers. There are two primary models of quantum computing. The first is gate-based quantum computing, which functions analogously to classical computing by using quantum gates to manipulate quantum information. This model facilitates the design of complex quantum circuits, which are the quantum counterparts to Boolean circuits in classical computation.

\noindent
In contrast, adiabatic quantum computing \cite{AL18} is based on the adiabatic theorem and, in principle, equivalent to quantum gate-based computing \cite{A+08}. In this model, the quantum system, which represents the search space of the given computational problem, gradually evolves from an initial, simple quantum state to a final state that encodes the problem's solution. The choice between these models depends on the nature of the problem being addressed and the availability in terms of hardware. %Gate-based systems are particularly effective for algorithmic computations, while adiabatic quantum computing is especially well-suited for optimization problems.

\subsection{Gate-Based Quantum Computing}

\noindent
%Regarding notation, we closely follow the textbook by Nielsen and Chuang.
The basic unit of information in classical information processing is a single bit that can exist in one of two states, represented as the integer numbers $0$ or $1$. Consequently, a sequence of \( n \) bits can represent \( 2^n \) unique values, with the bit register being in only one of these \( 2^n \) possible states. % at any given time.
A {\em quantum bit} (qubit) is the quantum analog of a bit and can assume the two basic states $\ket{0} = \icol{1\\0}$ and $\ket{1} = \icol{0\\1}$. 
%Single-qubit states are in 2-dimensional Hilbert space with orthonormal standard basis ($H^2$ = span($\ket{0}$,$\ket{1}$) $\cong \mathbb{C}^2$). 
Qubits adhere to the principles of quantum mechanics and can be implemented using various physical systems, such as the spin states of subatomic particles, ion traps, neutral atoms, or superconducting circuits.

\noindent
One of the intriguing properties of qubits is their ability to exist not just in the state \(|0\rangle\) or \(|1\rangle\) but in a {\em superposition} of both. An arbitrary {\em single-qubit state} can be expressed as:
\begin{align}\label{eq:single-qubit state}
    |\psi\rangle = \alpha |0\rangle + \beta |1\rangle = \begin{pmatrix}\alpha \\ \beta\end{pmatrix},    
\end{align}
where the coefficients \(\alpha\) and \(\beta\) are complex numbers that satisfy the normalization condition \(|\alpha|^2 + |\beta|^2 = 1\). A visual representation of a qubit is shown in Figure \ref{fig:block}.
When building {\em multi-qubit systems}, an \(n\)-qubit system provides access to a \(2^n\)-dimensional Hilbert space, where an arbitrary pure quantum state is defined as
\begin{align}\label{eq:multi-qubit state}
    |\psi\rangle = c_0 |0 \cdots 0\rangle + c_1 |0 \cdots 1\rangle + \cdots + c_{2^n-1} |1 \cdots 1\rangle,    
\end{align}
with \(c_i \in \mathbb{C}\) and $\sum_{i=0}^{2^n-1} |c_i|^2 = 1$.
The multi-qubit basis states, e.g., \(|0 \cdots 1\rangle = |0\rangle \otimes |0\rangle \otimes \cdots \otimes |0\rangle \otimes |1\rangle\), are tensor products of the individual qubits. The state \( |\psi\rangle \rightarrow [c_0, c_1, \cdots, c_{N-1}]^t \) possesses \(N = 2^n\) complex amplitudes, whose absolute squared values must sum to one. Thus, due to the principle of superposition, an \(n\)-qubit system is capable of encoding information that scales as \(2^n\), while classical systems are limited to \(n\).
A multi-qubit state is {\em entangled} if the states of its component qubits
cannot be described independently of each other as a tensor product. 
Quantum entanglement allows for non-local correlations such that operating on one 
component qubit changes the states of others instantaneously and independent of their separate locations; a prominent example are the entangled 2-qubit Bell states.

\noindent
In order for computation to be possible, there must be a way to manipulate quantum states. This is achieved by operators %acting on the Hilbert space \(H\). By definition, all operators 
that describe the evolution of a closed quantum system with unitary. In order to preserve the normalitation,  these have to be unitary, and therefore reversible, that is, for a {\em quantum operator} \(U\) must hold that \(U^\dagger U = I\). In fact, any operation on qubits can be described as a matrix operator. 
Notice that this is fundamentally different from classical computing, where operations are not required to be unitary and not even reversible.

\noindent
Another crucial difference between classical and quantum computation is how information can be accessed after processing. In classical computing, the state of each bit is well-defined and can be directly observed at any time, revealing whether it is in a state of $0$ or $1$. This observation does not alter the state of the bits, allowing us to access the exact information stored in the system without any disturbance. However, as mentioned above, in quantum systems the state of a qubit is generally a superposition of multiple states, meaning it can exist in a combination of both $0$ and $1$ simultaneously. To extract information from a quantum system, one must measure an observable physical quantity - which from now on we assume to be the binary number associated with the computational basis states. This {\em measurement process} forces the quantum system to \textit{collapse} from its superposition into one of the possible definite states. Crucially, this collapse is probabilistic, meaning that the outcome of the measurement cannot be precisely predicted but rather is governed by the amplitudes of the quantum state.  Furthermore, the measurement fundamentally alters the state of the system, making it impossible to retrieve the original superposition state after the measurement is performed. This is a key distinction from classical systems, where observation is non-intrusive and reversible.

\noindent
For the quantum state given by Eq. \eqref{eq:single-qubit state}, the measurement will yield one of the basis states 
\(|0\rangle\) or \(|1\rangle\) 
with classical output 0 or 1, respectively
%$|s\rangle$ where $s$ is the observed binary number with probability %$\left|c_s\right|^2$, thus the already mentioned normalization guarantees that there is a measurement output with certainty. 
The probability of observing \(|0\rangle\) is \(|\alpha|^2\), and the probability of observing \(|1\rangle\) is \(|\beta|^2\), with the total probability summing to 1 (\(|\alpha|^2 + |\beta|^2 = 1\)). After measurement, the qubit will be found in the state corresponding to the measurement outcome. 
For the state described in Eq. \eqref{eq:multi-qubit state}, the measurement outcome will be one of the \(2^n\) possible basis states. The probability of observing a particular multi-qubit state 
\(|\psi\rangle = |b_1 b_2 \cdots b_n\rangle\) is given by \(|c_{\text{index}}|^2\), where \text{index} represents the binary number corresponding to the combination \(b_1 b_2 \cdots b_n\), and the total probability is 1 (\(\sum_{i=0}^{2^n-1} |c_i|^2 = 1\)). After the measurement, the system collapses to the state observed, with the measurement outcome determining the final state of the system. Quantum state measurement is the only non-unitary (irreversible) operation in an ideal quantum computer, and according to the no-cloning theorem of quantum computing, in contrast to classical computing, it is not possible to generate an identical 
copy of an arbitrary quantum state. Note that the squaring of the amplitude over probabilities allows for negative and arbitrarily complex-valued amplitudes. These account for destructive interference which can be used in quantum alogortihms to supress undesired solution. This interference is a crucial distinction of quantum algorithms relative to classical stocahstic algorithms. Note also, that the need for the state vector to retain its norm (from which we infered that operations have to be unitary) can be understood as conservation of total probability.

\noindent
The aforementioned basic concepts of quantum computation are direct consequences of the postulates of quantum mechanics and provide a foundation for exploring how these principles can be applied in practice through {\em quantum circuits}, which function as the quantum counterpart of classical algorithms. 
%In fact, all unitary operations on qubits can be mathematically 
%described in Heisenberg's matrix form and implemented in 
%quantum gates.

\begin{figure}
    \centering
        \begin{quantikz}[column sep=8pt, row sep={30pt,between origins},slice label style={inner sep=5pt,anchor=south west,rotate=40}]  
        \lstick{$\ket{0}$}  &  \gate{H} & \qw & \gate{X} & \qw & \meter{}
        \end{quantikz}
            \caption{A simple quantum circuit demonstrating the combination of basic quantum gates. The circuit starts with a qubit in the \(\ket{0}\) state. The Hadamard gate (\(H\)) is applied first, creating a superposition of \(\ket{0}\) and \(\ket{1}\). Following this, a Pauli-X gate (\(X\)) is applied, flipping the qubit's state. The circuit concludes with a measurement, represented by the meter symbol, which collapses the qubit's state into either \(\ket{0}\) or \(\ket{1}\), producing a classical output.}
    \label{fig:quantum-circuit}
\end{figure}
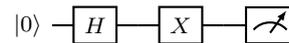

\noindent
A quantum circuit typically begins with qubits in a defined initial state, followed by the application of quantum gates that transform these states according to the intended computation. The circuit concludes with a measurement, which extracts classical information by collapsing the qubits into definite classical states.
Quantum circuit diagrams serve as a useful tool for visualizing quantum algorithms. Individual qubits are represented as horizontal lines, and the sequence of operations, also known as gates, is indicated by their position along these lines. An example of a simple quantum circuit is depicted in  Fig. \ref{fig:quantum-circuit}.

\noindent
Similar to classical computing there is a universal set of quantum gates, and 
quantum computing can simulate classical computing, while the converse 
is also true \cite{EW06}.  However, quantum computing solutions are in the 
complexity class of bounded-error quantum polynomial time for problems 
solvable in polynomial time by probabilistic quantum Turing machine with 
bounded error (error probability less or equal than 1/4). This class 
includes not only that of P but also interleaves with NP and PSPACE.
% ... siehe EW06 
Among other, that motivates research on quantum-supported problem-solving methods
beyond those with an already proven significant speed-up compared to 
their classical counterparts such as quantum prime factorisation \cite{Sho97} 
and quantum search \cite{PL24}. 

\begin{figure*}[ht]
    \centering
    \includegraphics[scale=1]{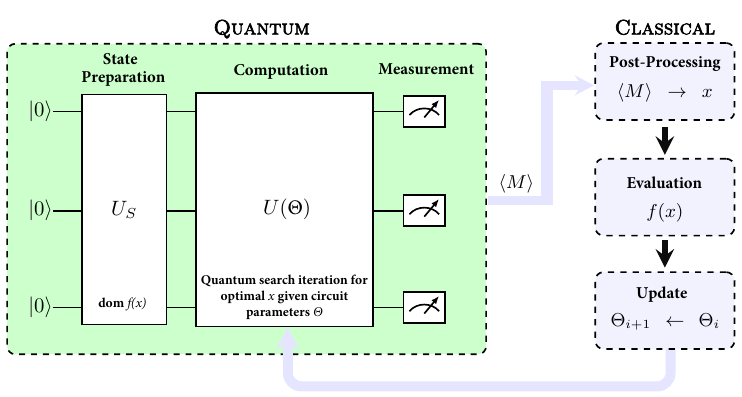}
    \caption{A Hybrid Quantum-Classical Optimization Workflow (adapted from \cite{Maca23+}). The diagram illustrates a hybrid quantum-classical optimization process. The quantum part involves three main stages: \textbf{State Preparation}: An initial quantum state is prepared using the unitary operator \( U_S \). \textbf{Computation}: The prepared state is processed by a parameterized quantum circuit \( U(\Theta) \) to search for the optimal solution \( x \) based on the parameters \( \Theta \). \textbf{Measurement}: The quantum state is measured to obtain the expectation value \( \langle M \rangle \). The classical part includes three steps: \textbf{Post-Processing}: The measured expectation value \( \langle M \rangle \) is processed to extract the classical variable \( x \). \textbf{Evaluation}: The function \( f(x) \) is evaluated based on the classical variable \( x \). \textbf{Update}: The parameters \( \Theta \) are updated using classical optimization algorithms to improve the search in the next iteration. The process iterates, with the updated parameters \( \Theta_{i+1} \) being fed back into the quantum circuit, forming a closed-loop optimization cycle between the quantum and classical computations.}
    \label{fig:hybrid}
\end{figure*}

\subsection{Adiabatic Quantum Computing}

\noindent
As an alternative, {\em adiabatic quantum computing} (AQC) \cite{AL18} performs problem-solving by gradually evolving a quantum system toward its lowest energy state as a solution. According to the adiabatic theorem, this gradual evolution allows the system to remain in its ground state throughout the process, theoretically enabling it to perform any quantum computation.
This process is designed to encode the solution to a given computational problem into this ground state, and AQC theoretically possesses the capability to perform any quantum computation, rendering it a universal approach to quantum computing. However, this theoretical potential relies on ideal conditions, including a perfectly isolated system, the ability to exert extremely precise control, and sufficiently long computation times to mitigate errors. Note that the total time for this evolution is controlled by the inverse of the energy gap, i.e., the difference between the two lowest energy eigenvalues during the sytem evolution, and an efficient adiabatic quantum algorithm is characterized by this gap shrinking only polynomially with problem size.

\noindent
A more practical application of AQC principles is {\em Quantum Annealing} (QA) \cite{KN98}, which involves guiding a quantum system towards a low-energy state but is specifically focused on finding approximate solutions to combinatorial optimization problems. 
%While QA is based on similar principles as AQC, it is tailored to address complex optimization tasks rather than providing a general-purpose quantum computational framework. 
Consequently, QA does not fully meet the criteria for universal quantum computing, making it more suitable for specific problem-solving scenarios rather than serving as a versatile computational tool akin to AQC.

\noindent
In quantum annealing, computational problems are formulated as {\em Quadratic Unconstrained Binary Optimization} (QUBO) problems, mathematically described as follows: 
\begin{align}\label{eq:QUBO}
    \text{minimize} \quad Q(\mathbf{x}) = \sum_{i,j} Q_{ij} x_i x_j
\end{align}
where \( \mathbf{x} \) is a vector of binary variables \( x_i \) (with \( x_i \in \{0,1\} \)), and \( Q_{ij} \) are the coefficients of the quadratic terms. The goal is to find the binary vector \( \mathbf{x} \) that minimizes the value of \( Q(\mathbf{x}) \).

\noindent
In practice, while QA does not guarantee the correct solution, it remains a powerful tool for addressing classically intractable problems. Notably, the QUBO formulation in Eq. \eqref{eq:QUBO} is NP-complete, implying that any NP-hard problem can be mapped into a QUBO formulation with only a polynomial overhead. This suggests that rather than viewing direct QA-based quantum algorithms 
%that solve given QUBO reformulated classical problems with quantum annealing 
as merely an alternative to classical or direct gate-based quantum computing for solving QUBO problems, it should be considered a tool that leverages quantum mechanics to provide potential advantages in specific cases, which warrant further exploration in practice.

\subsection{Hybrid Quantum-Classical Computation}

\noindent
While several theoretical results demonstrate that direct (gate-based or AQC-based) quantum algorithms can solve complex problems more efficiently than classical alternatives in terms of worst-case time complexity \cite{DJ92,Sho97,Aru+19}, these results invariably assume the existence of quantum computers with a very low logical error rates. This assumption necessitates the use of {\em quantum error correction} to safeguard quantum information against errors and noise that may arise during computation. Quantum computers are inherently sensitive to external disturbances, hardware imperfections, and spurious decoherence, 
%i.e., the tendency to become classical over time, 
all of which can introduce errors. 

\noindent
In contrast, NISQ devices, which represent the current state of quantum hardware, operate without full fault tolerance and have limited computational capabilities.
%In fact, current research focuses on leveraging NISQ devices, which may have an immediate impact on real-world applications. 
Such machines are not yet powerful enough to outperform classical computers and it is an open challenge to find clear quandaum advantage in NISQ. 

\noindent
Consequently, the concept of \textit{hybrid quantum-classical computation} has been proposed to exploit near-term quantum devices and benefit from the anticipated performance boost offered by quantum technologies.
Specifically, in the domain of gate-based quantum computing, {\em variational quantum algorithms} (VQAs) have been developed to address optimization problems by utilizing both classical and quantum resources. The quantum component, often referred to as a {\em variational quantum circuit} (VQC) or {\em parameterized quantum circuit} (PQC), plays a critical role in this hybrid approach.

\noindent
%In this framework, the classical input data $x$ of a VQA are initially pre-processed on a classical device to determine a normalized input quantum state for the PQC. Subsequently, a sequence of parameterized quantum gates $U(\theta)$, referred to as the {\em Ansatz}, is applied with randomly initialized parameters $\theta$. 
%In particular, a feature encoding unitary $U_\phi: \mathcal{X} \rightarrow \mathcal{F}$ maps the input vector $x \in \mathcal{X}$ to an $n$-qubit quantum state $\ket{\phi(x)} = U_\phi(x)\ket{0}$ in the Hilbert space $\mathcal{F}$ of $2^n \times 2^n$ Hermitian operators. 
%The quantum hardware then prepares this quantum state and computes $U(\theta)$ with randomly initialized parameters $\theta$. After multiple executions of $U(\theta)$, the classical component post-processes the measurements, an output $f(\theta)$, and generates an estimation of the target variable of interest. Finally, the parameters of the PQC are updated, and the whole cycle runs multiple times in a closed loop between the classical and quantum hardware.
\noindent
Within this framework, the classical input data \( x \) is initially pre-processed on a classical device to determine a normalized input quantum state for the PQC. Following this pre-processing, a VQA is executed, which comprises two sets of quantum gates: \( U_S \) and \( U(\theta) \).

\noindent
Quantum operations $U_S$ represent the data encoding (or embedding) step and consist of a sequence of quantum gates designed to generate a quantum state that represents the classical input $x$ as accurately as possible. The structure of $U_S$ is task-dependent, varying according to the specific computational problem being addressed. For instance, in combinatorial optimization problems, the encoding step involves generating a quantum state that represents the classical search space of the original problem. A prominent example is the Quantum Approximate Optimization Algorithm (QAOA) \cite{qaoa}, a well-known variational quantum algorithm for solving QUBO problems. In QAOA, the initial step of data encoding consists of generating a quantum state where the basis states encode all possible binary strings that might be solutions to the problem.
Conversely, in the context of machine learning, the encoding step involves mapping the set of features into a quantum state. This mapping can be achieved through various methods, such as using the basis states, amplitude encoding, or other encoding strategies. Each of these approaches has its own advantages and disadvantages. For a more detailed discussion of these methods and their implications, see \cite{Weigold21}.

\noindent
Subsequently, a sequence of parameterized quantum gates $U(\theta)$, referred to as the {\em Ansatz}, is applied with randomly initialized parameters $\theta$.
%The quantum hardware then prepares this quantum state and computes $U(\theta)$ with randomly initialized parameters $\theta$. 
After the execution of the PQC, which includes the quantum gates \( U_S \) and \( U(\theta) \), the result of the measurement 
%of the parametrized quantum state
is classically post-processed to obtain a classical output \( f(x) \). This output is then utilized to evaluate a task-dependent cost function. %, providing an estimation of the target variable of interest.
Based on this evaluation, the parameters of the PQC are updated using gradient descent or another optimization algorithm. This process is repeated iteratively in a closed loop between the classical and quantum hardware until the optimization converges or the desired performance is achieved.

\noindent
The strength of this approach lies in the adaptability of the architecture, which allows for customization through learning of the gate parameters in the PQC of the VQA to address various use cases. The entire procedure is depicted in Fig. \ref{fig:hybrid}. 

% \noindent
% It is important to note that how the encoding is performed depends on the specific computational task at hand. For example, in a combinatorial optimization problem, the encoding step involves generating a quantum state that represents the classical search space of the original problem. Conversely, for a machine learning task, the encoding step entails converting classical data into a quantum state \cite{Weigold21}.

\noindent
Quantum annealing can also be utilized within a hybrid framework. In this case, the approach involves designing hybrid quantum-classical algorithms that iteratively generate large QUBO problems, which can then be solved using quantum annealing. Here, only a specific computational component is delegated to quantum annealing, while the rest of the computation remains classical.

% principles of quantum computing and models 
% - gate-based, annealing, rydberg
% complexity quantum vs classical what is known
% in general: from classical problem to quantum solutions
% quantum conputer: status NISQ devices capabilities, physical for logical
% landscape? (is to some extent in the editorial service part)

% === 3 =====================================================
\section{Quantum Computing for AI}%
\label{sec:QC_for_AI}%
One direction of quantum AI refers to the use of quantum computing for solving computational problems in 
AI. As mentioned above, this concerns all subfields of AI such as automated planning, machine learning, 
computer vision, natural language processing, and multi-agent systems, for each of 
which we provide selected results and insights in this section. The other direction of QAI, namely, 
AI for quantum computing is covered in Section 4.

% For only some - dont have much time anymore !! - topic field of QAI:QC4AI 
% ** summary of only few main results and applications 
%    (derived from just a few selected paper results 
%      plus references for further information on this topic field 
%           for the interested reader) 
% ** only few open questions
%
% So, no real subsections as in a real survey anymore 
% - and some topic field blended out - but 
% a sequence of some short (not enough for a subsection) 
% and some comparatively longer text paragraphs on topic fields.

% --------------------------------------------------------
% Quantum Machine Learning 

\subsection{Quantum Machine Learning}

\noindent
{\em Quantum machine learning} (QML) seeks to harness the principles of quantum computing to perform traditional machine learning (ML) tasks \cite{B+17,B+22,B+20}. Although theoretical research indicates that fault-tolerant quantum computing could accelerate the training of various ML algorithms 
%by speeding up typical linear algebra operations and 
providing a computational advantage in terms of worst-case time complexity \cite{Robe14,Maca20,Maca23}, current quantum hardware is not yet powerful enough to implement these algorithms effectively. % at large scale. 
Consequently, in QML the focus has shifted towards leveraging hybrid quantum-classical computation with VQAs (cf. Section 2.3), which aims to exploit gate-based near-term quantum devices to develop innovative models and potentially achieve performance gains from quantum technologies.

%\noindent
%As mentioned above, variational quantum algorithms are particularly designed to tackle optimization problems using both classical and quantum resources. 
%The input data $x$ of a VQA are initially pre-processed on a classical device to determine a normalized input quantum state for the parameterized quantum circuit. 
%A feature encoding unitary $U_\phi: \mathcal{X} \rightarrow \mathcal{F}$ maps the input vector $x \in \mathcal{X}$ to an $n$-qubit quantum state $\ket{\phi(x)} = U_\phi(x)\ket{0}$ in the Hilbert space $\mathcal{F}$ of $2^n \times 2^n$ Hermitian operators. The quantum hardware (the PQC) then prepares this quantum state and computes $U(\theta)$ with randomly initialized parameters $\theta$. After multiple executions of $U(\theta)$, the classical component post-processes the measurements and generates an estimation of the target variable of interest. Finally, the parameters of the PQC are updated, and the whole cycle runs multiple times in a closed loop between the classical and quantum hardware.
%The strength of this approach lies in the adaptability of the architecture for different use cases through the {\em learning} of the gate parameters.

\noindent
The hybrid QML approach with the use of VQAs (cf. Section 2.3) shares significant similarities with the training of classical neural networks (NNs), notably in their reliance on parameterized models, gradient-based optimization techniques, and structured layers for approximating complex functions. 
As such, this common ground has paved the way for the development of {\em quantum neural networks} (QNNs), which essentially are the use of PQCs for machine learning applications. 
QNNs are widely used in both supervised and reinforcement learning. In {\em quantum supervised learning} \cite{M24}, these algorithms typically involve fitting a parameterized function to a training dataset, 
%The data is pre-processed and encoded into a quantum state, and a PQC is employed to process this quantum state. 
%Notably, the PQC 
that can represent complex hypothesis functions that classical models might struggle with. %The primary distinction between QNNs based on hybrid computation and classical neural networks lies in the execution of function calls to $f(x; \theta)$: in the former, a PQC is employed, while in the latter, a classical neural model is used. This subtle yet fundamental difference aims to leverage the unique capabilities of quantum computing, potentially offering advantages in processing efficiency and learning capabilities for specific tasks.
Therefore, the use of QNNs in supervised learning seeks to identify problem classes that are intractable for classical approaches from a learning perspective, rather than expedite the training process in terms of wors-time complexity.

\noindent
Similarly, QNNs can be effectively utilized in {\em quantum reinforcement learning} (QRL) \cite{QRL24,DTB17}. In reinforcement learning (RL), an agent interacts with an environment to maximize cumulative rewards by learning optimal policies. Hybrid quantum-classical approaches of RL can 
employ QNNs to encode and process states and actions within the environment, leveraging a potential quantum advantage in representing and exploring complex state-action spaces. QNNs can be used to approximate value functions or policy distributions, while classical 
components handle the optimization of these quantum parameters through iterative updates. This hybrid strategy can enhance the learning efficiency and representational power of RL algorithms, potentially solving problems that are challenging for classical methods alone.

\noindent
Recently, a quantum deep reinforcement learning method has been developed and experimentally evaluated on noisy gate-based quantum simulation for the use case of safe navigation by self-driving cars that do not need a quantum device on-board during testing but training only \cite{SMK23}. The QRL method leverages the model capacity of quantum neural networks, only requires a few dozen noisy qubits and relies on hybrid quantum-classical computations to be effective. The experimental evaluation provided evidence in favor of quantum utility in terms of faster and more stable training with fewer parameters compared to the classical counterpart. These results suggest that existing noisy quantum computing devices with a few tens of qubits might soon become viable alternatives to overcome the challenges faced by classical methods in, for example, enhancing autonomous systems and optimizing large distribution networks through quantum-supported reinforcement learning.

\noindent
The focus on using quantum algorithms for unsupervised learning primarily involves fault-tolerant quantum approaches that can theoretically provide a computational speedup in terms of worst-time complexity. Key advancements include the introduction of quantum algorithms that utilize amplification techniques in clustering problems \cite{ABG13,DH96}, the proposal of a quantum k-means and k-nearest centroid algorithms \cite{LMR13,Kere19}, and recent efforts focused on graph sparsification achieving by leveraging superposition access to classically stored graph weights \cite{AdW20}.

% QUANTUM ML subfields ....

    % 1 - supervised
    
    % 2 - unsupervised
    
    % 3 - reinforcement - single QRL
    % QRL-survey \cite{QRL24} \cite{DTB17} 
    % model-based offline QRL \cite{E+24})

% multiagent RL - only mention it in principle 
% - see refs Gate-based: DS24, PK+23, YK+23 & QA-based: N+20, KT+23
% some text for it is in the QMAS paragraph, and scheduling re application.

% --------------------------------------------------------
% Quantum Planning and Scheduling

\subsection{Quantum Planning and Scheduling}

\noindent
{\em Quantum automated planning and scheduling} (QPS) research focuses 
on quantum-supported means of automated planning (QP) and scheduling (QS) in AI, and vice versa. 
Automated planning methods in AI can be divided into online and offline planning 
each with certainty or under uncertainty. 
Offline planning is decoupled from the subsequent execution of produced 
plans and gets no feedback about it during planning, while online planning 
is interleaved with a controlled action execution in a closed-loop manner. 
In many real-world applications, action planning under uncertainty is 
required that allows for actions with non-deterministic effects and 
incomplete initial states caused by only partial observability of the 
environment such as in partially observable Markov decision processes 
(POMDP). Depending on the chosen technique of planning with certainty or under 
uncertainty, a plan can take the form of, for example, a finite sequence of 
primitive actions, a conditional action plan, or an action policy on belief 
states with maximal expected utility. 
For a more comprehensive introduction to automated planning, we refer the 
interested reader to, for example, \cite{G+16}. \\
\noindent
In any case, automated planning methods in AI are known to be computationally 
expensive, such as classical (state-space or planning graph-based) action 
planning already being PSPACE-complete, while belief states in POMDPs
and non-determinism as well as interactive POMDPs make it even worse 
exponentially. Nevertheless, automated planning of symbolic AI is 
considered more explainable in general than deep learning-based POMDP solutions,
and approximated online POMDP planning methods are successfully used in many 
practical applications such as robot navigation. 
The fundamental question therefore arises as to whether, to what extent and 
under which conditions quantum-supported AI planning, particularly for (interactive) POMDPs, 
is feasible and may lead to a significant reduction of planning time and space compared 
to the classical counterparts. As of today, not much is known in this regard yet.  \\
\noindent
For example, a classical POMDP models an agent acting in a partially 
observable stochastic environment. A first quantum-based POMDP model was 
proposed in \cite{BBA14}, which is intended to generalize POMDPs and has 
the same complexity for the strategy (plan) existence problem, i.e. is PSPACE-hard 
for a polynomial (undecidable for infinite) time horizon. In this model, 
actions and observation process are represented by a super-operator on 
quantum-encoded environment states assumed by the agent. 
However, it is unclear to what extent this model is useful for 
quantum-supported planning and learning agents in POMDPs in concrete terms. \\
\noindent
In \cite{C+16}, a quantum MDP model (QMDP) is defined through a potential 
energy function of the quantum system under consideration. It is shown how  
model-based learning by approximate value iteration for POMDPs can be applied 
to such a model. However, how QMDPs are modeled when the energy function is 
unknown, and whether this model may serve as a basis for quantum-supported 
approximated interactive POMDP planning is not known either. 
Another open question is to what extent an adaptation of hybrid neuro-symbolic 
methods for learning-assisted planning in POMDPs \cite{PK19,DCH23} 
for quantum computing would be feasible and beneficial. \\
\noindent
In \cite{C+22}, a quantum-supported method for optimal path planning in robot navigation
is presented that leverages Grover's quantum search in a classical tree-search procedure.
It is simulated for gate-based quantum computation and shown theoretically to always 
provide a quantum speedup up to that of the Grover algorithm \cite{grover}.
Earlier work \cite{N+06} proposes a QP method that adapts and benefits from 
standard quantum search for planning in MDPs using Dynamic Programming (DP)
and a heuristic for controlling a discrete time quantum walk on the MDP state graph,
which reduces the number of states visited during a DP iteration compared to classical.

\noindent
{\em Quantum scheduling} (QS) research actually focuses on the use of quantum computation
for solving the NP-complete problem of job shop scheduling (JSS) and its modern variants 
that allow for more flexibility (FJSS) in the context of Industry 4.0. 
Roughly, the FJSS problem is to find a feasible schedule that assigns constrained job 
operations to multi-purpose machines with, for example, a minimal production makespan 
in total. There exist many different kinds of variants of this problem and classical 
approximate solution methods based on genetic algorithms, artificial neural networks, 
particle swarm optimization and reinforcement learning \cite{Z+19,L+18}. 
In agent-based solutions for FJSS, individual machines, carriers, jobs or individual 
operations are often represented by agents in a respective multi-agent system. \\
\noindent
However, there are only very few quantum-supported solution methods for some 
variants of the FJSS problem yet. These QS methods are limited to quantum 
genetic algorithms and quantum PSO, in which iterative evolutionary sets of 
suitably quantum-encoded individuals can lead to approximate optimal 
solutions, for example by means of quantum rotations \cite{ZH19,CYH19,SX19}. \\
\noindent
In \cite{AHY20,AAY22}, an iterative quantum-supported hybrid MILP (Mixed 
Integer Linear Programming)-based optimal solution of the JSS problem is 
decomposed in a way that is suitable for hybrid quantum-classical computing. 
The hybrid method solves a relaxed MILP problem of the original JSS problem 
with a classical MILP solver to find possible optimal assignments of jobs to 
machines. This is followed by solving related sequencing QUBO problems 
that correspond to individual machines with a D-Wave quantum annealer to
search for feasible schedules based on those assignments. The reported 
experimental results show a significant computational time speedup of this 
method over the classical solver Gurobi for JSS problem instances comprising 
of up to 280 machines and jobs in a specific scheduling dataset.\\
\noindent
Multi-agent reinforcement learning (MARL) is one classical solution approach 
for FJSS with dynamic changes of optimization conditions and configurations 
\cite{LPT23,IS+23,PH+19}. Currently, there are only very few initial, feasible approaches 
of quantum-supported MARL (QMARL) for this specific problem class \cite{PK+23,YK+23} 
but without any experimental analysis and concrete insights on their benefits compared 
to classical counterparts yet. \\
\noindent
Another NP-hard problem in the AI domain of automated planning and scheduling is the 
bin-packing problem, which is to find the minimum number of bins of fixed capacity 
required to pack a set of items of varying size without exceeding the bin capacities. 
Recently, a first QS method has been presented that solves this problem reformulated as 
a QUBO problem on a quantum annealer \cite{CML24}. The method utilizes the Augmented 
Lagrangian method to account for the bin packing constraints and heuristic penalty 
multipliers, scales with increasing problem size but does not outperform the selected 
classical counterpart in runtime yet due to current limitations of quantum annealing 
hardware. \\

\noindent
It is apparent that more in-depth theoretical and experimental investigations are needed
on which types of computationally hard AI planning may benefit from the adoption 
of direct or hybrid quantum computation under which conditions and assumptions of integration
compared to classical solutions in general, and for which practical use cases in particular. 
Research in the other direction of QPS, that is, the use of AI planning and scheduling 
for advanced manufacturing and operating quantum devices, is as relevant as the use of
AI for the same purpose regarding classical computing devices but actually occurs less 
prominent in the literature (cf. Sect. 4).

% --------------------------------------------------------
% Quantum Vision

\subsection{Quantum Computer Vision}

\noindent
{\em Quantum computer vision} (QCV) research is mainly concerned with the investigation of the 
feasability and benefits of quantum-supported computer vision methods for the perception of 
intelligent agents in AI. Research in the other direction of QCV, that is, the usage of computer 
vision methods for advances in the building and operating quantum computing devices 
actually appears rather neglected. \\
\noindent
The QCV subfield of QAI formed in the early 2000s \cite{VB03} and attracted renewed attention
with recent advances in quantum image processing. The latter requires the representation and 
processing of a given digital image on a quantum computer, and the final conversion of the 
processed quantum image into a classical image. Quantum image representations such as qubit 
lattices and normal arbitrary quantum superposition states store color image information using 
amplitudes, phases, or basis quantum states \cite{LP23,Li+19}. 
For the processing of quantum represented images, there are already quite a few,
mostly quantum annealing-based methods available for tasks such as image recognition and 
classification \cite{C+20,U+24,K+24}, image synthesis \cite{F+24}, object tracking and detection 
\cite{LG20}, graph matching \cite{S+20}, as well as for motion and image segmentation 
\cite{A+22,V+23,V+24}.\\ 
\noindent
For example, the problem of unsupervised graph-based image segmentation 
is (a) to construct a weighted undirected graph from a given image with set of vertices (pixels), 
set of edges (synergies between pixels), and set of weights (similarity between pixels),
and then (2) to find the best partition into disjoint subsets such that the sum of weights 
between different subsets is minimized. 
This NP-hard problem has been recently solved in \cite{V+23} with a quantum-supported method that
first reformulates segmentation as a graph-cut optimization problem, maps it into the topology of 
and runs it on the considered quantum annealer, in this case, a D-Wave Advantage, then retrieves the 
results of quantum measurement and eventually generates the segmentation mask from them.
Despite the shared and remote access of the D-Wave device, comparative experimental 
evaluation results revealed that this QCV method outperformed the classical solver Gurobi 
on the same task in terms of runtime with only slightly sub-optimal solution quality. 
In this regard, it is particularly valuable when collecting labeled data is costly and speed is of 
essence. In \cite{V+24}, quantum algorithms for the same purpose 
are analyzed that allow to scale the number of qubits exponentially with respect to the 
input size to use current gate-based quantum computing devices.\\
\noindent
Motion segmentation, on the other hand, aims to detect independent motions in two or several 
input images. The QCV method presented in \cite{A+22} solves this problem reformulated as 
a QUBO problem directly on a D-Wave quantum annealer with reported on-par performance compared to 
classical solutions. \\
\noindent
Recently, in \cite{U+24}, several variations of a quantum hybrid vision transformer were
analyzed for solving an image classification problem in high-energy physics with the result of, again, 
an on-par performance compared to classical counterparts with a similar number of model parameters.
For small-scale medical image datasets, it has been shown in \cite{K+24} that 
quantum transformer models with quantum attention layers may perform better than classical
vision transformers for this purpose in terms of asymptotic run time and fewer model parameters. \\
\noindent
In \cite{F+24}, two quantum hybrid diffusion models for image synthesis are presented.
The first model replaces convolutional ResNet layers with hybrid quantum-classical 
variational quantum circuits only at the vertex, while the second additionally does so 
in the second block of the encoder part. The experimental evaluation via simulation in PennyLane
indicate that such models are of benefit in the sense that they synthesize better-quality images 
and converge faster with a lower number of parameters to train.

% --------------------------------------------------------
% Quantum Natural Language Processing

\subsection{Quantum Natural Language Processing}

% examines the prospects of quantum computing in natural language processing %(NLP), including tasks like language translation, sentiment analysis, and %text summarization.

\noindent
{\em Quantum natural language processing} (QNLP) is concerned with the 
representation and processing of natural language through quantum computational means.
The other direction of QNLP, that is the utilization of NLP means for quantum computing tasks 
remains fully unexplored yet.
As summarized in \cite{GPE22}, most QNLP works leverage quantum superposition to model uncertainties 
and ambiguity in language or entanglement to describe both composition and distribution of syntax 
and semantics effectively. The current QNLP methods are mainly developed for the NLP tasks of question 
answering, text classification and translation. In \cite{CFMT22}, Coecke and colleagues show that 
QNLP is not just the quantum counterpart to NLP but allows to combine linguistic structure and 
semantics or meanings in one quantum computational system, in fact, represent this kind of knowledge 
in respectively composed variational quantum circuits more efficiently and inherently than in the 
classical case. 
The common basis of most QNLP approaches is the Categorical Distributional Compositional (DisCoCat) 
diagram model for natural language \cite{CSC10} with which one can encode the meaning of words and 
phrases as quantum states and processes, hence as quantum circuits. In this regard, the implicit 
flows of meanings in respective diagrammatic reasoning due to the underlying structure are exposed 
rather than encoded in black boxes of neural transformer-based large language models (LLM) in 
classical NLP. \\
\noindent
The potential speedup of running the DisCoCat model on quantum hardware was discussed first in 
\cite{ZC16}, while in \cite{W24,L+23} the potential of quantum computing techniques such as quantum 
search and quantum neural networks for faster training and testing of LLMs, and vice versa, is 
conceptually discussed but not yet demonstrated in practice. In this context, complementary work 
includes quantum-inspired approaches such as the recently reported compression of LLMs with 
quantum-inspired tensor networks by supposedly more than 80\% without compromising 
accuracy \cite{L24}. \\
\noindent
However, more advanced experimental insights on and development of hybrid quantum-classical QNLP 
methods, in particular quantum neural networks, for NLP tasks are needed but limited by current NISQ 
quantum computing hardware. For example, as in QCV, the potential of quantum-supported transformers 
compared to classical transformer neural network architectures might be worth to investigate. 
For more information on QNLP in Quantum AI, we refer the interested reader to, 
for example, \cite{W24,GPE22,CFMT22}

% --------------------------------------------------------
% Quantum Multi-Agent Systems}%
\subsection{Quantum Agents and Multi-Agent Systems}

\noindent
{\em Quantum multi-agent systems} (QMAS) research mainly focuses on two strains: (1) development of autonomous agents and multi-agent systems for hybrid quantum-classical computational environments, and 
(2) quantum-supported methods for coordination and cooperation in multi-agent systems. 
In modern AI, the concept of an intelligent agent and multi-agent system is at the core 
\cite{RN16,W99} and integral part of many agent-based scientific, industrial and commercial 
AI applications. 
Roughly said, in a multi-agent system, multiple homogeneous or heterogeneous 
agents coordinate their actions to achieve joint goals and carry out tasks 
flexibly, autonomously and interactively in complex environments with 
cooperative or competitive settings depending on the considered application problem. \\
\noindent
There are quite some tools and frameworks for the engineering of autonomous agents 
and multi-agent systems for environments of classical computing 
\cite{ajan,jade,jadex,jason,agentspeak}. 
However, architectures and approaches to build and operate agents with quantum computational 
capabilities in AI applications to run on one or multiple networked quantum computers or in
hybrid quantum-classical computing environments are very rare. 
In fact, concepts, tools and frameworks for quantum multi-agent system programming based on the 
currently available quantum programming frameworks and quantum computational models are needed 
but still missing and topics of future research in QMAS.\\
\noindent
Early work in this direction \cite{K03,KS07} proposed a first conceptual classification and 
architecture of quantum multi-agent systems. 
More recent, a quantum modeling approach for reactive agents with subsumption architecture has been 
proposed and exemplified for a simple ball picking robot in \cite{K+21}. The agent state is encoded 
as the superposition of the tensor products of a 4-qubit perception vector with a 5-qubit action 
vector, and the simple quantum control circuit for the robot is shown for gate-based quantum 
computation but not yet evaluated.  \\
\noindent
Research on quantum-supported means of coordination and cooperation in multi-agent systems 
is still in its infancy. In \cite{N08}, two first quantum-supported methods for coordination in 
a quantum multi-agent system are presented for gate-based quantum computing devices.
These methods were concerned with quantum versions of Kernel-based coalition negotiation 
and a specific contract net protocol with embedded auction.  
The used quantum coalition protocol provides a marginal speedup in computation and quadratic 
reduction in communication between agents compared to the classical counterpart, 
while the quantum contract net version offers no computational speedup but more data privacy to 
the bidding agents.\\
\noindent
Likewise, \cite{C+07} initially discusses the potential of leveraging entangled quantum states for 
coordination in (mixed) multi-agent systems by means of a quantum public-goods protocol and quantum 
auctions for resource allocation. Their experiments on the latter indicates, for example, 
that the quantum version may provide more privacy than the classical (first price) auction but 
at the cost of lower economic efficiency.\\
\noindent
More recently, quantum multi-agent reinforcement learning (QMARL) for adaptive coordination in 
collaborative settings for the quantum gate-based model \cite{DS24,PK+23,YK+23} or quantum 
annealing \cite{N+20,KT+23} gained some interest. 
For example, in \cite{DS24} a QMARL version for centralized training and decentralized execution 
(CTDE) relies on distributed advantage actor-critic architecture with a quantum critic uniquely 
spread across the agents and coupling of local observation encoders through entangled input qubits 
over a quantum channel. That eliminates explicit sharing of local observations of agents and 
reduces classical communication overhead. Whereas in \cite{PK+23,YK+23} alternative QMARL versions 
for CTDE are proposed with a quantum critic at a central server and agents sending their local 
observations via a classical channel.\\
\noindent
In any case, experimental evidence in favor of, or against some quantum utility compared to 
classical MARL methods is shown for simple toy domains only \cite{DS24}, if at all. 
More investigations in this regard are definitely required to assess the potential of QMARL.\\
\noindent
One prominent class of micro-level coordination techniques for multi-agent systems \cite{OM06} 
in competitive environments is coalition formation based on cooperative game theory. In general, 
coalition formation refers to situations in which groups of individually rational agents intend 
to work jointly in (temporary) coalitions with binding agreements in order to accomplish their 
tasks they cannot accomplish individually \cite{E+13,SCD22}. Research in this area focuses on two 
main problems, namely, coalition negotiation and coalition structure generation \cite{R+15,R07}. 
Roughly said, for a given coalition game $(A, v)$ with a set $A$ of rational agents and 
characteristic or coalition value function $v$, the goal is either (a) to negotiate a 
game-theoretically stable coalition structure as partition of $A$ with individual payoffs for 
coalition members, or (b) to generate the optimal coalition structure that maximizes the social 
welfare or joint profit of the multi-agent system as a whole without individual payoff distribution. 
Both problems are computationally expensive, thus potential candidates for the investigation of 
quantum-supported solution alternatives. 
For coalition negotiation this holds subject to the chosen stability criteria like the exponential 
Kernel and Shapley value, while coalition structure generation (CSG) is a NP-hard optimization 
problem.\\
\noindent
Since early work on quantum (non-)cooperative games \cite{IT02,I05,N08,CIA12,I+23,D+23}, 
only recently methods for quantum coalition structure generation have been developed and 
experimentally evaluated against classical counterparts revealing evidence in favor 
of some quantum utility \cite{VMK23a,VMK23,VMK23b}. 
As an example, \cite{VMK23b} presents a direct quantum solution of the CSG problem reformulated 
as a QUBO problem for both quantum gate-based simulation and quantum annealing devices, 
while \cite{VMK23a} proposes a hybrid quantum-classical CSG method for induced subgraph 
games. In particular, the latter leverages quantum annealing to iteratively solve the embedded NP-hard 
optimal splitting problem in only linear runtime to find the best bipartition of agents by moving down 
partition hierarchy until no value-increasing bipartitions remain. In fact, it explores a larger 
portion of the solution space compared to other approximate classical bottom-up solvers and 
outperforms them with its quadratic runtime in the number of agents and an expected worst-case 
approximation ratio of 92\% on standard benchmarks.\\
\noindent
Remarkably, quantum coalition formation methods have already been applied to practical 
use cases of energy management \cite{eon1,eon2}. The management of energy resources presents 
complex computational challenges, particularly with scenarios involving dynamic energy consumption and 
optimizing energy distribution in micro-grids and larce-scale power networks as well. 
For example, the use case problem considered in \cite{eon2} is to find an optimal coalition structure 
of energy prosumers with the aim to minimize network management costs through optimized power flow 
analysis, that is to maximize the overall network efficiency. The results of using the quantum CSG
method from \cite{VMK23a} for solving this problem on quantum annealers showcase quantum utility 
compared to classical standard solutions. Advancing in this direction will facilitate the broader 
adoption of quantum AI solutions for real-world energy management challenges. \\
\noindent
Another use case of quantum CSG methods in the aero-space domain was recently investigated in 
\cite{V+24a}. The problem to find constellations of large-scale low-earth satellite networks 
such as StarLink networks by dynamic clustering of moving LEO satellites with minimal 
inter- and intra-cluster communication is NP-hard. According to initial experimental results,
the used quantum CSG method on a quantum annealer outperformed the classical state-of-the-art
solver for this problem on real-world orbital datasets from Celestrak related to Starlink satellites.\\
\noindent
Coalition formation methods have also been leveraged in the domain of transport and logistics 
for collaborative solutions of the NP-hard problem of vehicle routing and variants of the 
capacitive vehicle routing problem (CVRP) \cite{MX+23,ZHG16,WM+17}. However, quantum-supported 
solutions for the same are still rare and topic of future research \cite{F+19,I+19}. 
For example, in \cite{F+19} a CVRP is solved hybrid quantum-classical through 
classical clustering and optimal route planning in each cluster, while the latter is done 
with a direct quantum optimization algorithm for solving the corresponding TSP(Traveling Salesman Problem)-QUBO problem 
on a D-Wave quantum annealer. The method did not provide a clear benefit in solution quality 
and runtime, though the latter might change with the advent of a more advanced quantum annealer 
in the future.

% ========================================================
\section{AI for Quantum Computing}%
\begin{figure}
    \centering
    \includegraphics[width=0.8\columnwidth]{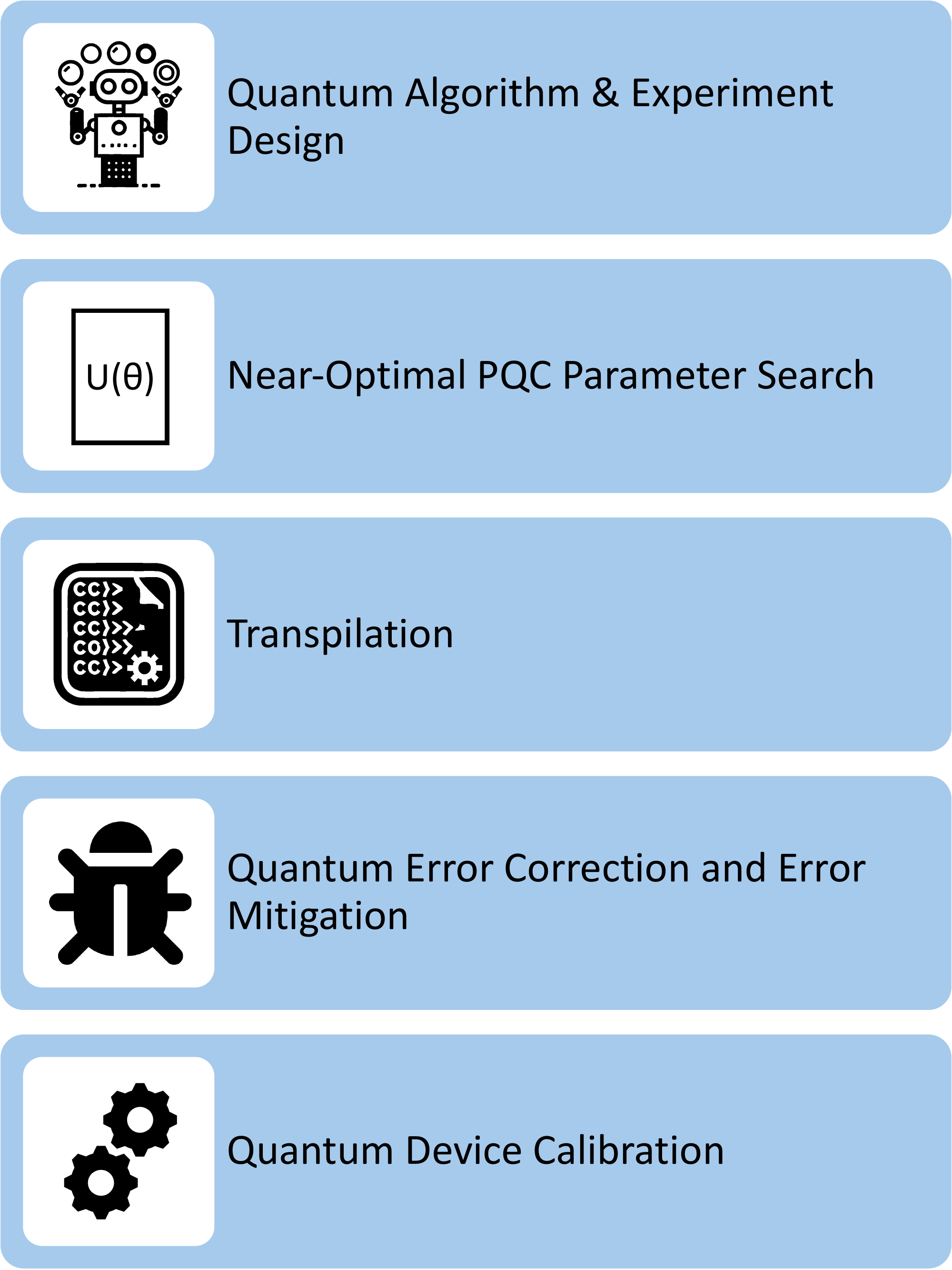}
    %width=\linewidth]{sections/Image_AI4QC.pdf}
    \caption{Stack of tasks in the building and operation of quantum computing 
              devices for which AI techniques (currently mainly from ML) are 
              utilized.}
    \label{fig:AI4QC}
\end{figure}

\label{sec:AI_for_QC}%
In this section, we turn our attention to the other direction of QAI, namely, the use of AI for quantum computing. It summarizes selected approaches to exploit AI, particularly ML methods, in support of the whole process from quantum algorithm and experiment design, search for near-optimal parameters, transpilation of quantum circuits, error correction during execution, and the calibration and design of quantum devices as shown in Fig. \ref{fig:AI4QC}. It can be seen as a particular example of the field of AI for science. 
%The current trend in this research reflects a significant shift towards ML-assisted problem-solving
%methodologies, driven by the limitations of human intuition in quantum algorithm development and experimental design, as well as the challenges posed by (NP-)hard problems in other domains where classical computational resources are already insufficient.

% Eigentlich bräuchten wir fehler-tolerante QC, um die Probleme der NISQ devices zu lösen und damit Fehler-tolerantes QC zu ermöglichen ;)
% at each step ai can assist, currently predominantly ml approaches are utilized,
% overall we currently see a shift from non-ai algorithms to ai algorithms espacially with QISKITs newest alpha/beta software -> first approaches are tested at greater scale right now
% similar to the introduction of ai in classical computing we are with quantum computing at a stage, where we have a lot of NP-hard problems to solve, where we hope to achieve a big step forward by utilizing ai
% give an overview over the used algorithms
% we still have a lot of the same problems as we have in classical computing and since it reached it limits we are shifting our focus to ml as it has helped us also progress in classical computing

% ... habe entsprechende Überschriften der subsections eingefügt

%\subsection{Quantum Algorithm Design and Implementation}
%\noindent
%Though the design and implementation of quantum algorithms in general 
%is manually done by human experts, there are quite a few approaches to 
%leverage ML methods in support of the experiment design and protocol 
%development. %for quantum communication and quantum cryptography. 

\subsection{Quantum Algorithm and Experiment Design}
\label{subsec:AI4QCExp}

\noindent
Though the design and implementation of quantum algorithms in general 
is manually done by human experts, there are quite a few approaches to 
leverage ML methods in support of both the experiment design and protocol 
development. There is a common agreement that humans might not be 
predestined to design new experiments and protocols in quantum computing 
when recent experimental findings deliver counter-intuitive results. 
Most approaches to experiment design suggest the use of reinforcement learning 
and minimizing the influence of the human developer. \\
\noindent
In this regard, the framework \textsc{Melvin} \cite{KM+15} offers the use of ML methods to discover new experimental methods for creating and manipulating complex quantum states. 
%\textsc{Melvin} is able to learn quantum phenomena and solve specific problems by optimizing 
%quantum building blocks, and learns from simpler systems to speed up the discovery of more complex experiments. In particular, 
It creates experiments with optical components, arranges them randomly, calculates and analyzes 
quantum states, and simplifies configurations based on user-defined criteria, if the desired 
properties are met. Remarkably, \textsc{Melvin} produced correct and useful but unfamiliar, 
asymmetric techniques that are hard to understand intuitively.\\
\noindent
Inspired by \textsc{Melvin}, \cite{MP+17} and \cite{WM+20} use the physics-oriented ML approach of projective simulation based on RL for autonomous experiment design. For example, the authors in \cite{MP+17} applied projective simulation for developing complex photonic quantum experiments with multi-photon states entangled in high dimensions. As a result, the approach unexpectedly rediscovered advanced experimental techniques, 
%suggesting that machines could play a significant creative role in research. 
in particular by interacting with a simulated optical table using a set of optical elements to create experiments. It places elements on the table, analyzes the resulting quantum states, and receives rewards based on the given task and state before iterating the process.
\\
\noindent
In \cite{WM+20} the same approach is used to discover and optimize protocols in quantum communication over long distances. 
%They investigate whether RL can independently identify and enhance crucial quantum protocols. Again, 
Their RL agent successfully rediscovered established protocols such as for quantum teleportation, entanglement purification, and quantum repeaters. The results suggest that RL could be effective in finding solutions that outperform human-designed ones for long-distance communication challenges, especially in scenarios with asymmetric conditions.
%, such as nonuniform channel noise and varying segment distances. 
%Their agent was able to find schemes that outperform human-designed ones primarily in asymmetric situations. 
%The generation of unknown experiments depends on then only on the formulation of the task.
\\
\noindent
%In the area of modern optical communication systems, the
%need for ML to enhance both classical and quantum communication was 
%recently emphasized in \cite{ZM+20}. Their study highlights key insights 
%on the design of Raman amplifiers and phase tracking at the quantum 
%limit. For example, ML techniques are particularly effective in learning 
%complex input-output mappings, which facilitates system optimization, 
%the development of signaling and detection schemes for intricate or 
%analytically intractable channels, as well as ultra-sensitive signal 
%detection. By applying multi-layer neural networks for inverse system 
%design of ultra-wideband Raman and SDM amplifiers, and Bayesian inference 
%for precise optical phase tracking, the authors have achieved significant 
%improvements in quantum-noise limited bit error rate performance.
%\\
%\noindent
%Instead of using a reinforcement learning algorihtm, Costa et al. \cite{CO+21} investigate how classical machine learning algorithms can enhance the precision of quantum phase estimation using differential evolution (DE) and particle swarm optimization (PSO) to fine-tune feedback policies aimed at reducing Holevo variance amidst different types of noise. 
In contrast, \cite{CO+21} investigates the use of differential evolution and particle swarm optimization for improving the precision of multi-particle entanglement-free quantum phase estimation. Both methods are leveraged to fine-tune feedback policies aimed at reducing different types of noise. In fact, both methods showed superior robustness and precision compared to non-adaptive methods, particularly for scenarios where noisy, non-entangled qubits serve as sensors in quantum sensing and metrology. 
% In particular, both DE and PSO methods adhered to 
% the Standard Quantum Limit (SQL) curve for phase estimation, with DE slightly outperforming PSO. Specifically, DE achieved a Holevo variance scaling of approximately $N^{-0.75}$, while PSO achieved $N^{-0.64}$. Both algorithms demonstrated superior robustness and precision compared to non-adaptive methods, particularly under higher noise conditions. The protocols can be applied to various experimental platforms, including optical Mach–Zehnder interferometry, superconducting qubits, trapped ions, and NV centers in diamond.
% In fact, both methods showed superior robustness and precision compared to non-adaptive methods, particularly under higher noise conditions. 
% The protocols can be applied to various experimental platforms, including optical Mach–Zehnder interferometry, superconducting qubits, trapped ions, and NV centers in diamond.
\\
\noindent
% XAI 
% maybe tell about EU AI Act
Related research on explainable AI (XAI) for quantum computing is concerned with the use of XAI techniques not only to explain the designed quantum algorithms but  quantum circuits in general and PQCs for QML (cf. Section 3.1) in particular. 
%As pointed out earlier understanding quantum computing experiments and developing an intuition seems impossible. Therefore, in the following we discuss several approaches, where eXplainable AI (XAI) will help us understand critical components of qunatum circuits in general and Parameterized Quantum Circuits (PQCs) in particular.
For example, \cite{SS+22} reports that PQCs introduce probabilistic errors due to quantum measurements, which complicates the use of traditional XAI methods. Moreover, with the phase space of a quantum circuit expanding exponentially with the number of qubits, executing XAI methods in polynomial time becomes challenging. The authors evaluated the adaptation of the XAI methods IG (Integrated Gradients) and baseline SHAP (SHapley Additive exPlanations) for PQCs using truncated Fourier series. 
%More concrete, their method 
%PolynomialSHAP computes SHAP values efficiently by approximating PQC 
%outputs with multivariate polynomials, while their method qSHAP samples 
%quantum circuits to perform Fourier decomposition and then uses 
%PolynomialSHAP for computing SHAP values. 
Based on their findings, the authors introduced qSHAP (quantum SHAP), which scales with the number of features rather than qubits, making it suitable for larger quantum circuits, and shows more robustness against noise compared to the other two approaches.\\
\noindent
In \cite{HG+23}, SHAP is investigated in the form of quantum Shapley (QShap) values to assess the impact of single or group of quantum gates in PQCs, similar to evaluating feature importance in classical ML. QShap values are model-agnostic within quantum domains, treating gates as players in a coalition game to measure their contribution to tasks like expressibility, entanglement capability, or classification quality (see also \cite{SJA19}). To handle the uncertainty in quantum computing, QShap values are based on uncertain Shapley values, which account for measurement noise and decoherence. The authors tested QShap values on quantum support vector machines (QSVM), quantum generative adversarial networks (QGANs) and QAOA (cf. Section 2.1), reporting on the significance of individual gates in these algorithms.
\\
\noindent
In contrast, using the inception or so-called deep dreaming method, \cite{JA+23} analyzes a neural networks' understanding of quantum optics experiments. 
%For this purpose, they invert the network to observe its "dreams about quantum systems", revealing insights into its internal logic. 
%Quantum experiments are modeled as colored, weighted multi-graphs with photon paths as vertices and correlations as edges. The focus lays on four-qubit, two-dimensional quantum systems, with the network discovers optimal quantum graphs through deep dreaming. 
This method shows the ability of the used neural network to find novel configurations beyond the initial data and reveals a progression from simple to complex feature recognition in the layers of the network.
\\
\noindent
Overall, most progress is made in the area of experiment design with algorithm design lacking behind. As to XAI, first results, primarily based on Shapley values, were presented. Further, XAI will get extremely important in QML as the EU AI Act defines transparency rules, besides others, for all applications of AI, including QAI. Further, a better understanding of the influence of individual gates, as proposed in \cite{HG+23}, might help in optimizing the depth and width of PQCs during the transpilation process.
% designing optimal Pulse sequences might be also a good idea

\subsection{Near-Optimal PQC Parameter Search}
\noindent
\textit{Warm-starting} of quantum algorithms refers to the process of finding near-optimal parameters of PQCs in hybrid quantum-classical algorithms without executing them. They aim to reduce the number of quantum circuit evaluations required during the optimization. Classical warm-starting protocols rely on relaxations of the problem which have to be solved classically to find a good initial state \cite{EMW20}. In the following, we summarize selected approaches that involve classical ML algorithms for this purpose.\\
\noindent
For example, \cite{AAG20} investigates four regression-based machine learning methods to speedup the QAOA optimization loop. These methods include Gaussian process regression, linear model, regularized support vector machine, and regression tree. Of these, Gaussian process regression achieved the best results. In particular, despite using a relatively small training set, this approach generalized well and
%($66$ instances for training and $264$ for testing)
%to new instances beyond the training set
led to an average reduction of $44.9\%$ in the number of optimization iterations across all local optimization procedures.\\
%The reduction in runtime was especially significant for higher target depth implementations. For example, the Nelder-Mead optimizer saw a decrease in function calls of $12.3\%$ for a target depth of $p=2$, $43.3\%$ for $p=3$, and $57.7\%$ for $p=5$.
\noindent
In \cite{JC+21} and \cite{LL+24}, the integration of graph neural networks (GNNs) with QAOA is explored to improve the parameter initialization. 
%The authors of \cite{LL+24} focused on solving the Max-Cut problem with
Various GNN architectures such as graph convolutional networks, graph attention networks, graph isomorphism networks, and GraphSAGE  \cite{HYL17} were investigated in \cite{LL+24}. Experimental evaluation with benchmarks for their performance in initializing QAOA showed that all GNNs provided a more stable and reliable initialization compared to random initialization.
%Experimental results reported in \cite{JC+21} revealed that the GNN method shows performance comparable to that of the SDP relaxation method but with significantly faster inference times, which makes it scalable to larger problems. 
One of the strengths of GNN-based initialization is its ability to generalize across different graph sizes. This means that a GNN trained on smaller graph instances can still perform well on larger instances, providing a flexible and efficient warm-starting mechanism for QAOA. Including graphs of different sizes the GNN method in \cite{JC+21} generated solutions of approximately $95\%$ of the quality of the Goemans-Williamson algorithm. %Including graphs of different sizes in the training set 
%increased the performance ratio to approximately $95\%$. 
\\
\noindent
The ability of matrix product states (MPS) for improving QML algorithms is explored in \cite{DB+21} by addressing the issue of vanishing gradients, or barren plateaus that complicate the training of PQCs. The authors propose optimizing MPS with classical methods such as density matrix renormalization group and time-evolving block decimation to approximate solutions for quantum circuits. The optimized MPS is converted into unitary matrices for quantum circuits using decompositions. %such as KAK. 
Experiments showed that MPS-initialized circuits for problems like Max-Cut on a six-vertex graph and image classification with Fashion-MNIST achieved better performance, that is, with fewer gradient steps and faster convergence compared to random or identity matrix initialization.
\\
\noindent
Overall, ML algorithms show promise in the warm-starting of variational quantum algorithms. Future research might aim at methods to find the optimal parameters of the PQCs of such algorithms directly using some sort of ML algorithms and therefore replacing the classical optimization loop completely. 
In a similar direction, future ML-based algorithms might also assist with choosing the best hyper-parameters such as for the Ansatz circuit and the number of layers, which was not explictly mentioned by the papers referenced in this section.

\subsection{Transpilation of Quantum Circuits}
\noindent
Given the challenges of NISQ devices, such as low fidelity and short coherence times, quantum circuits must be designed with minimal gates. Afterwards, quantum circuit transpilation includes the decomposition of non-native gates into native ones and the addition of SWAP-gates for connectivity compatibility, followed by an optimization to minimize resource usage. 
%Quantum circuit transpilation involves modifying a quantum circuit to optimize it for a specific quantum machine. 
IBM Qiskit \cite{qiskit} divides this quantum circuit transpilation process into six stages: Init (prepares the circuit by unrolling instructions and validating them), layout (maps virtual to physical qubits considering connectivity and calibration), routing (ensures backend compatibility with additional gates), translation (converts gates to the backend’s basis set), optimization (reduces circuit complexity), and scheduling (applies hardware-aware scheduling). Currently, ML-based 
algorithms are utilized for most of the steps to find optimal transpiled circuits.
These include evolutionary algorithms, deep neural networks, and RL methods.

\noindent
\textit{Evolutionary algorithms} including genetic algorithms, genetic programming, ant colony optimization, and evolutionary deep neural networks are used in all stages of the transpilation process \cite{KSU21}. Further, \cite{DCL24} and \cite{FGL22} utilize evolutionary algorithms such as evolutionary deep learning for an initial qubit mapping.
%Dahi et al. \cite{DCL24} have created their own dataset with data collected from four IBM machines over a 10-month period and introduced their own deep learning model based on evolutionary deep neural networks for circuit mapping. In \cite{FGL22} evolutionary algorithms are also used for an initial qubit mapping. Afterwards, reinforcement learning is used, which is present later in this section.
\\
\noindent
\textit{Neural networks} are primarily used for circuit mapping and circuit optimization. 
In \cite{G21}, neural networks are utilized to significantly speed up the circuit mapping process while maintaining mapping accuracy and reducing the required computational resources. The experiments showed this on a 5-qubit IBM Q processors and compared to classical state-of-the-art mapping algorithms over a special data set for training \cite{GDat21}. However, \cite{DCL24} argue that the method and dataset above lacks correctness, generalization, and diversity of the dataset. 
\\
\noindent
In \cite{Me22}, on the other hand, a framework based on long-short term memory (LSTM) networks is proposed to decide whether a quantum circuit can be optimized in the first place. The reasoning is that, as mentioned above, the optimization process is a complex task and it is not guaranteed to receive a more resource efficient quantum circuit. Thus, if their framework predicts that the quantum circuit cannot be optimized, classical computing resources are saved. Results show that their ML model is able to decide if an arbitrary quantum circuit can be optimized above a certain threshold with an accuracy of $96.7\%$. 
\\
\noindent
Similar to \cite{Me22}, Quetschlich et al. introduce in \cite{TUM22a} and \cite{TUM23} an algorithm, which finds the optimal compilation flow for a given quantum circuit. \cite{TUM23} builds on their previous work \cite{TUM22a} by shifting from reinforcement learning to a simpler, scalable supervised learning approach for predicting optimal compilation options. The proposed tool recommends the best compiler options for quantum circuits, targeting end-users who may struggle with option selection. Using a statistical classifier, the tool predicts the best technology, device and compiler settings with accurate results for about $75\%$
of unseen test circuits. 
\\
\noindent
Paler et al. \cite{PL+20} introduce a quantum circuit mapping heuristic, named  QXX, and its ML-enhanced version QXX-MLP to improve layout, routing, and optimization stages. The goal is to map circuit qubits to physical qubits while optimizing the circuit depth. QXX-MLP uses a multi-layer perceptron to infer optimal parameter values for QXX, reducing circuit depth by a Gaussian function estimation. QXX achieves depth ratios about $30\%$ lower than Qiskit on shallow circuits and performs competitively with Qiskit and TKET on deeper circuits. Beyond that QXX-MLP achieves almost instantaneous layout performance, significantly reducing the quantum circuit compilation time.
\\
\noindent
\textit{Reinforcement learning} is primarily utilized to find near-optimal gate synthesis for a specific hardware back-end.
A deep reinforcement learning method for approximating single-qubit unitaries is proposed in \cite{MP+21}. This method aims to reduce the overall execution time by learning a general strategy through a single pre-compilation procedure. The authors highlight the trade-off between the length of the sequence and execution time, suggesting that their approach can potentially allow for real-time operations by improving on this trade-off.
\\
\noindent
In \cite{FGL22}, the problem of quantum circuit placement \cite{MFM08}
is addressed as a bilevel optimization problem to minimize SWAP counts. The authors use a deep reinforcement learning algorithm for the lower-level optimization, improving SWAP costs through state space encoding. For the upper level, as mentioned earlier, an evolutionary algorithm is used. This ML-based framework reduces SWAP gates by up to $100\%$ and runtime costs by up to $40$ times compared to heuristic methods. 
\\
\noindent
Fosel et al. \cite{FN+21} describe a deep RL agent that optimizes quantum circuits tailored to specific hardware, improving efficiency for near-term devices. Their method reduced average circuit depth by $27\%$ and gate count by $15\%$ in 12-qubit random circuits, demonstrating effective resource reduction. The RL agent also scales well, optimizing larger circuits effectively. 
\\
\noindent
Recently, Kremer et al. \cite{KV+24} presented an RL-based method for synthesizing quantum circuits, including Clifford, Linear Function, and Permutation circuits, which directly matches native device instructions and constraints. This approach eliminates the need for extra transpilation steps, optimizing the transpiling process. It also enhances circuit routing, reducing the two-qubit gate depths and counts more effectively than heuristics like SABRE. That led  to a better performance on quantum devices with up to $133$ qubits. For 8- to 10-qubit quantum volume \cite{P+22} circuits, the RL method achieved a reduction around $20\%$ in the CNOT depth.
\\
\noindent
In contrast to the aforementioned ML algorithms, the authors of \cite{T+24} discuss the effectiveness of classical automated reasoning and formal methods in AI, namely, decision diagrams, SAT solvers, and graphical calculus-based methods such as the ZX-calculus, for the compilation of quantum circuits.
The considered compilation tasks are classical simulation, optimization, synthesis, and equivalence checking of quantum circuits. As one result, the authors expect that the use of automated reasoning methods 
may play a role in other quantum computing applications as well, such as for the finding of ground states, phase transitions, and quantum error correction.
\\
\noindent
Overall, we have seen several ML algorithms targeting one ore more transpilation stages. We are currently at a stage, where first ML-enhanced transpilers are rolled out to the end-users for testing in IBM Qiskit \cite{KV+24}. However, as long as we have only access to NISQ-devices we are always striving to find shallower circuits. Furthermore, besides reducing the depth of quantum circuits, also reducing the width, that is, the number of qubits required, is another topic of future research in this context.
%ing, which was not explicitly mentioned in the referenced papers of this section.
% quantum circuit optimization is the most important topic, as it helps us to utilize current NISQ devices much better

\subsection{Quantum Error Correction and Mitigation}
\noindent
{\em Quantum error correction} and {\em error mitigation} are both relevant processes on our way to fault-tolerant quantum computing. While error correction, as the name suggests, actively corrects errors during the execution of a quantum circuit, error mitigation targets the readout error and therefore is a classical post-processing step. % Problem of the overhead of quantum error mitigation
\\
\noindent
A significant challenge with quantum error mitigation is the necessity for a large number of circuits, which have to be run in advance. In essence, each possible bitstring has to be tested for read-out errors. In \cite{LW+23}, ML is used to improve quantum error mitigation by predicting near noise-free values from noisy quantum output. Their key innovation is that trained ML models can mitigate errors without additional circuits, reducing overhead compared to zero-noise extrapolation. The authors explore several models, including linear regression, random forests, multi-layer perceptrons, and graph neural networks, finding that all except graph neural networks outperform traditional methods, with random forests consistently performing best. In fact, ML-based quantum error mitigation reduces the quantum resource overhead by $30\%$ and the runtime overhead by $50\%$ compared to zero-noise extrapolation. 
%The RF model trained on $500$ circuits and tested on $2,500$ circuits showed a superior error reduction for Trotter circuits in a transverse-field Ising model, with significantly lower execution errors than ZNE.
\\
\noindent
Quantum error correction is actively researched as it will deliver fault-tolerant quantum computing. In \cite{CL+22}, an ML algorithm for continuous quantum error correction is proposed.
%, which outperforms discrete methods by avoiding the need for entangling gates and ancilla qubits. 
The approach facilitates recurrent neural networks to identify bit-flip errors in continuous noisy syndrome measurements. 
%RNNs are trained on synthetic signals that simulate real-world imperfections such as auto-correlation, transient dynamics post-bit-flip, and steady-state drift. 
%Their RNN-based CQEC algorithm is compared with a traditional double threshold scheme and a discrete Bayesian classifier, showing superior performance across tasks and achieving fidelity comparable to the Bayesian model. Furthermore, it significantly suppresses the logical error rate compared to the bare qubit or uncorrected three-qubit system.
\\
\noindent
Further, in \cite{HGU23} a deep neural network decoder for quantum error correction on IBM quantum processors was developed and benchmarked.
%focusing on the heavy-hexagonal architecture. By adapting the heavy-hexagonal code to fit IBM's qubit layout, 
The study demonstrates the DNN decoder's capability to efficiently process syndrome data and correct errors, outperforming the traditional Minimum-Weight-Perfect-Matching method in certain aspects. The performance of the DNN decoder was validated through simulations and experiments on IBM devices, showing promise for real-time, scalable error correction, which is crucial for fault-tolerant quantum computing.
\\
\noindent
Overall, the research on ML algorithms for quantum error correction and quantum error mitigation is still in its early stages. In addition, the use of 
AI methods other than from ML for quantum error correction remains to be investigated.
% Problem with real-time error correction, as a high amount of data has to be processed

\subsection{Calibration of Quantum Computing Devices}
\noindent
The {\em calibration and design of quantum computing devices} such as superconducting or ion trap gate-based quantum computers, quantum annealers and quantum sensors can benefit from classical ML.
\noindent
This is originally an application of the field of quantum optimal control theory \cite{GBC+15,KBC+22}, which aims at finding the right analogue controls for given hardware to perform desired quantum tasks. This field has already used mathematical tools similar to those of AI and is now rephrasing a lot of its work in modern AI language \cite{WRP+21,RRM+23,YAS+24}. In fact, this has recently led to the tongue-in-cheek remark statement, that AI is in fact a subfield of quantum control theory. 

These methods typically have a model-based and an experiment-based part and ideally also use experimental data to improve the model, in order to create a digital twin and provide a shortcut for optimal control. This is done in isolation e.g. in \cite{SKW16,CSG+05,HFW20,GFH+17,SCL+24,GSG+23} and integrated in the commercial offerings of a number of startups \cite{RPWM22,BAH+21,MLK+20}.

For example, \cite{NSP21} presents a practical, efficient, and model-independent ML method for Bayesian parameter estimation (BPE) in quantum systems. Traditional BPE methods often require extensive calibration and explicit modeling of the measurement apparatus, which makes them impractical for complex systems. The authors frame parameter estimation as a classification problem solved using supervised learning techniques, where the output of the neural network is a Bayesian posterior distribution centered at the true parameter value, bounded by Fisher information. This approach requires fewer calibration measurements and is model-independent outperforming conventional calibration-based BPE.
%, particularly with limited data and for non-Gaussian states, using a multipartite entangled non-Gaussian state of 50 qubits. It is also robust to noise and experimental imperfections, making it practical for real-world applications.
\\
\noindent
%Wozniakowski et al. \cite{WT+20} present a ML approach that leverages prior scientific knowledge to improve model generalizability, especially in data-scarce scenarios with an existing domain model. 
In \cite{WT+20}, the energy spectrum of a Hamiltonian on a superconducting quantum device is predicted, outperforming the current state-of-the-art by over $20\%$. The method uses multi-target regression to predict multiple related variables, uncovering relationships between them by employing explainable AI techniques. This approach significantly improves the accuracy of quantum device calibration. 
%, which is crucial for quantum simulations and experiments. 
%The authors also employ explainable AI techniques to interpret results and reveal parameter dependencies, providing insights into the quantum device's behavior.
\\
\noindent
To simplify the calibration process, \cite{CG+19} introduces a method to calibrate quantum photonic sensors using neural networks.
%in handling large datasets and 
The approach relies on data collected using available probe states, reducing overhead, and implicitly accounting for imperfections. %Experimental validation with a quantum phase sensor based on N00N states showed that the method achieves uncertainties close to the theoretical Cramér-Rao bound, with specific uncertainties of \(F_{1000} = 1.25\), \(F_{5000} = 1.39\), \(F_{10000} = 1.48\), and \(F_{40000} = 1.53\) for different event counts. 
The neural network demonstrated robustness to noise and scalability, making it suitable for future quantum technologies. 
%Issues like boundary effects near specific phase values were noted, suggesting improvements in the training range. This method significantly advances the practical and efficient calibration of quantum sensors, integrating classical machine learning techniques with quantum computing.
%
Quantum hardware design can be made more systematic and be improved by AI as well, as shown, for example, in \cite{GCH+24,MHG+21}.

\noindent
Overall, the calibration of quantum devices can be practically improved through the use of ML. In conjunction with ML-assisted experiment design (see Sec. \ref{subsec:AI4QCExp}) one even might find advantages in the use of ML for designing new quantum computers with better fidelities
%$T_1$ and $T_2$ coherence times, was ist das? -> T_1 decoherecne time 0 to 1 and T_2 dephasing time + to - and vice versa
and connectivity.

%
%siehe z.b. auch: 
%https://www.quantentechnologien.de/forschung/foerderung/weitere-projekte/ai4qt.html

%e.g. \cite{C+23,KLFM24}

% ========================================================
\section{Conclusions}%

\noindent
Research in the interdisciplinary and nascent field of quantum AI is concerned with the use
of quantum computing for addressing computationally
hard problems in AI, and vice versa. 
So far, an impressive progress was made in both directions of QAI research. In fact, there are quite a few quantum-supported solutions of selected hard optimization problems in AI with different degrees of potential quantum utility 
for relevant use cases in various domains such as manufacturing, automated driving, finance, and energy management. 
In addition, initial research revealed that quantum computing itself may benefit from the use of ML for optimizing the control, performance and calibration of quantum computational devices. 

\noindent
In our view, future QAI research across all its subfields should focus even more on investigations under what conditions and settings in concrete 
(industrial) use cases are direct or hybrid quantum-classical solutions feasible with what quantum utility in practice. Further, the appropriate and timely transition to and investigation of the feasibility and potential of nowadays QAI methods on future built non-NISQ devices is another challenge. 
That is particularly important in the context of current expectations of the economic value of QAI applications in relevant industries. Among other, this requires both the physics and computer science communities to even more join forces, and on the other hand, a further, sustainable support of research on both quantum AI and the building of more resourceful, fault-tolerant quantum computing devices at government and industry level worldwide.

\begin{acknowledgements}
We gratefully acknowledge support of this work 
by the German Ministry for Education and Research (BMB+F) 
under project grant Q(AI)2.
\end{acknowledgements}


\section*{Declarations}

{\bf Conflict of interest} The authors declare that they have no conflict of interest.\\

\noindent
{\bf Open Access} ~This article is licensed under a Creative Commons Attribution
4.0 International License, which permits use, sharing, adaptation,
distribution and reproduction in any medium or format, as long
as you give appropriate credit to the original author(s) and the source,
provide a link to the Creative Commons licence, and indicate if changes
were made. The images or other third party material in this article are
included in the article's Creative Commons licence, unless indicated
otherwise in a credit line to the material. If material is not included in
the article's Creative Commons licence and your intended use is not
permitted by statutory regulation or exceeds the permitted use, you will
need to obtain permission directly from the copyright holder. To view a
copy of this licence, visit \url{http://creativecommons.org/licenses/by/4.0/}.

%{\bf Funding} ~Open Access funding enabled and organized by Projekt DEAL.
%



\end{document}